\documentclass{vietnam}
\usepackage{hyperref}
\usepackage{amsmath,amssymb}

\def\etal{{\it et al.}}
\def\unit#1{\ {\rm #1}}

\def\gagamma{g_{a\gamma}}

\def\be{\begin{equation}}
\def\ee{\end{equation}}
\def\bea{\begin{eqnarray}}
\def\eea{\end{eqnarray}}

\newcommand{\Photo}{}

\begin{document}
\vspace*{4cm}
\title{Searches for ultralight vector and axion dark matter with KAGRA}

\author{Yuta~Michimura$^1$,
Takumi~Fujimori$^{2}$,
Hiroki~Fujimoto$^{3}$,
Tomohiro~Fujita$^{4,1}$,
Kentaro~Komori$^{3,1}$,
Jun'ya~Kume$^{5,6,1}$,
Yusuke~Manita$^{7}$,
Soichiro~Morisaki$^{8}$,
Koji~Nagano$^{9,10}$,
Atsushi~Nishizawa$^{11,12,1}$,
Ippei~Obata$^{13}$,
Yuka~Oshima$^{3}$,
Hinata~Takidera$^{3}$}

\address{$^1$Research Center for the Early Universe, University of Tokyo, Bunkyo, Tokyo 113-0033, Japan\\
$^2$Department of Physics, Osaka Metropolitan University, Sumiyoshi, Osaka 558-8585, Japan\\
$^3$Department of Physics, University of Tokyo, Bunkyo, Tokyo 113-0033, Japan\\
$^4$Department of Physics, Ochanomizu University, Bunkyo, Tokyo 112-8610, Japan\\
$^5$Dipartimento di Fisica e Astronomia ``G. Galilei'', Universit\`a degli Studi di Padova, via Marzolo 8, I-35131 Padova, Italy\\
$^6$INFN, Sezione di Padova, via Marzolo 8, I-35131 Padova, Italy\\
$^7$Department of Physics, Kyoto University, Sakyo, Kyoto 606-8502, Japan\\
$^8$Institute for Cosmic Ray Research, University of Tokyo, Kashiwa, Chiba 277-8582, Japan\\
$^9$LQUOM, Inc., Yokohama, Kanagawa 240-8501, Japan\\
$^{10}$Institute of Multidisciplinary Sciences, Yokohama National University, Yokohama, Kanagawa 240–8501, Japan\\
$^{11}$Physics Program, Graduate School of Advanced Science and Engineering, Hiroshima University, Higashi-Hiroshima, Hiroshima 739-8526, Japan\\
$^{12}$Astrophysical Science Center, Hiroshima University, Higashi-Hiroshima, Hiroshima 739-8526, Japan
$^{13}$Kavli IPMU, University of Tokyo, Kashiwa, Chiba 277-8583, Japan}

\maketitle
\abstracts{We have proposed using laser interferometric gravitational wave detectors to search for ultralight vector and axion dark matter. Vector dark matter can be probed through oscillating forces on suspended mirrors, while axion dark matter can be detected via oscillating polarization rotation of laser beams. This paper reviews these searches with the KAGRA detector in Japan, including the first vector dark matter search with KAGRA's 2020 data and installation of polarization optics for axion dark matter search during the upcoming 2025 observing run.}

\section{Introduction}
The first direct detection of gravitational waves in 2015 marked the dawn of gravitational wave physics and astronomy~\cite{GW150914}. Since then, the global network of gravitational wave detectors, including LIGO, Virgo, and KAGRA, has reported the observation of over 250 events. These detections have revealed new perspectives on the universe and have enriched our understanding of astrophysical phenomena.

The extraordinary sensitivity of gravitational wave detectors also enables the search for other types of fundamental physics, such as dark matter that alters the interference fringes. In particular, laser interferometers are well-suited for detecting ultralight dark matter, which produces oscillatory changes in the interference fringes at a frequency of $f \simeq 242 \unit{Hz} \times (m_{\rm DM}/10^{-12} \unit{eV})$.

Several innovative approaches have recently been proposed, with some having already been successfully demonstrated. Scalar dark matter, which induces time variations in the fine structure constant or particle masses, can be probed by measuring size changes in mirrors, and the first such search was conducted using the data from the GEO600 interferometer~\cite{GEOScalar}. These fields can also be searched for by measuring the acceleration caused by the spatial gradient of mirror masses~\cite{MorisakiScalar}, with upper limits derived from the LIGO's third observing run~\cite{FukusumiScalar}.

Axion-like particles are another candidate that can be explored by detecting oscillating polarization rotation of laser beams due to axion-photon coupling~\cite{DANCE,ADAM-GD,ADAM-GD2}. While gravitational wave detectors have not yet been used for axion dark matter searches, initial results from tabletop interferometric experiments such as DANCE~\cite{DANCEOshimaProc,DANCEFujimotoProc,DANCEFirst}, LIDA~\cite{LIDAFirst} and ADBC~\cite{ADBCFirst} have been reported.

Vector dark matter weakly coupled to the standard model sector can be searched for by measuring oscillatory forces acting on mirrors~\cite{MichimuraVector,MorisakiFinite}. Searches have already been conducted using data from the third LIGO-Virgo observing run~\cite{O3Vector}, the KAGRA O3GK run~\cite{O3GKVector}, and the LISA Pathfinder mission~\cite{AndrewLPF,FrerickLPF}. Additionally, gravitational wave detectors can be used to probe spin-2 dark matter, as it produces signals similar to those of gravitational waves~\cite{ManitaSpin2,ManitaExploring}.

In this paper, we present the current status of vector and axion dark matter searches using KAGRA detector in Japan~\cite{MichimuraProc}. We adopt the natural unit system with $\hbar=c=\epsilon_0=1$.

\begin{figure}
\begin{minipage}{0.49\linewidth}
\centerline{\includegraphics[width=\linewidth]{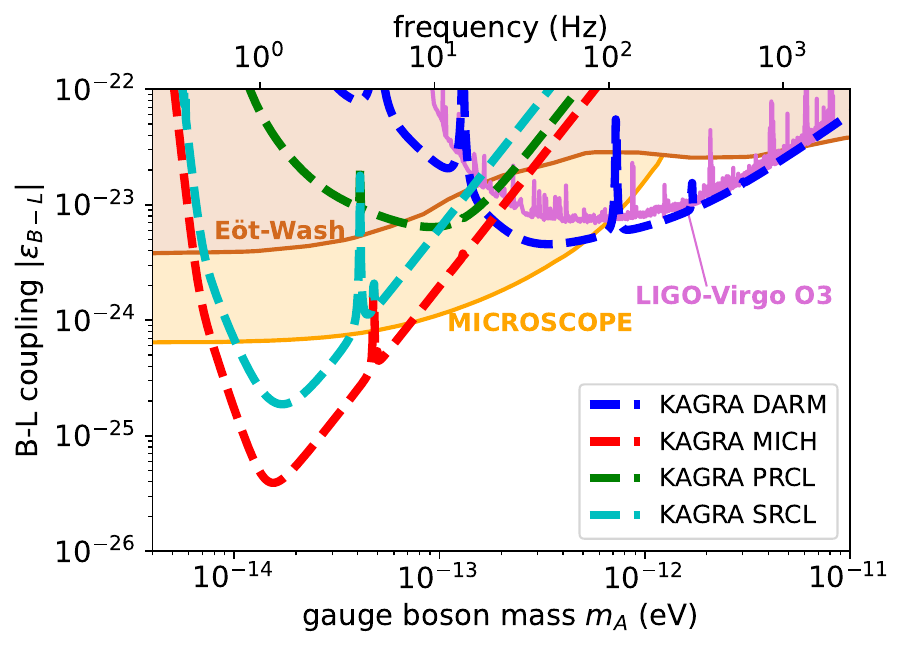}}
\end{minipage}
\hfill
\begin{minipage}{0.49\linewidth}
\centerline{\includegraphics[width=\linewidth]{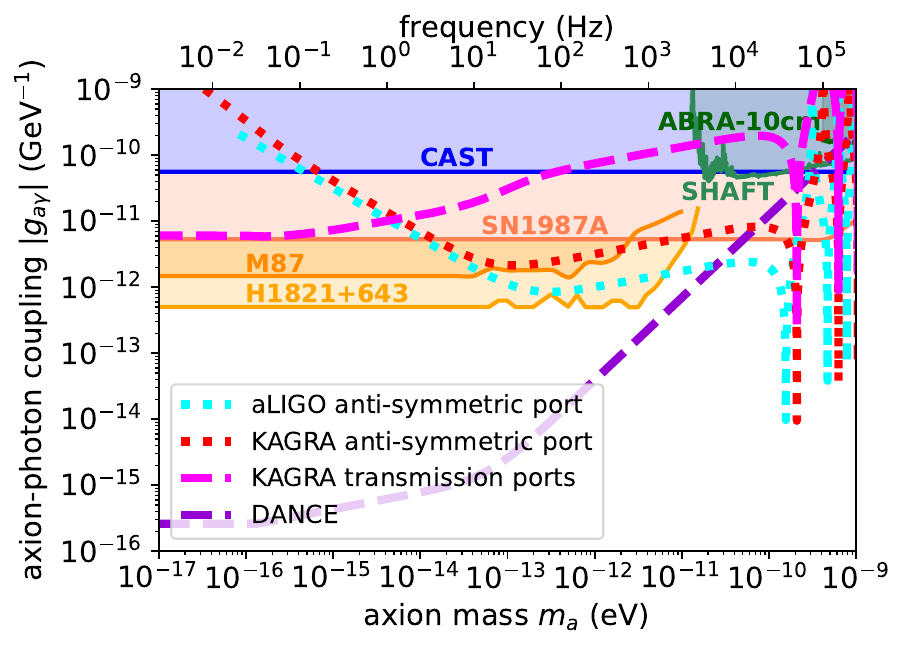}}
\end{minipage}
\caption[]{The projected sensitivity of KAGRA for $B-L$ vector dark matter~\cite{MichimuraVector} (left) and axion dark matter~\cite{ADAM-GD,ADAM-GD2} (right) with the measurement time of 1 year. The shaded regions show limits from fifth-fore searches with E\"{o}t-Wash torsion pendulum~\cite{EW2008,EW2012} and MICROSCOPE satellite~\cite{MICROSCOPE2018} and LIGO and Virgo's third observing run~\cite{O3Vector} (left), and limits from CAST~\cite{CAST}, SHAFT~\cite{SHAFT}, ABRACADABRA-10cm~\cite{ABRA} experiments, astrophysical bounds from the gamma-ray observations of SN1987A~\cite{SN1987A} and the X-ray observations of M87 galaxy~\cite{M87} and H1821+643 quasar~\cite{H1821+643} (right). The projected axion sensitivity for DANCE~\cite{DANCE} and LIGO~\cite{ADAM-GD2} are also shown for comparison.}
\label{fig:sensitivity}
\end{figure}

\section{Initial results from the vector dark matter search using KAGRA}
The existence of dark matter suggests new physics beyond the standard model. One particularly appealing framework is the $B-L$ (baryon minus lepton) extension of the standard model, which provides a natural explanation for the small neutrino masses through the seesaw mechanism and addresses the matter–antimatter asymmetry via leptogenesis. In the standard model, $B-L$ is conserved and anomaly-free, and the $U(1)_{B-L}$ symmetry can be gauged without requiring additional components. If the gauge boson associated with this new $U(1)_{B-L}$ symmetry acts as ultralight dark matter, it induces an oscillating force on the suspended mirrors of a laser interferometer, leading to detectable length changes.

Under a vector dark matter field, expressed as $\vec{A}(t,\vec{x})=A_0 \vec{e}_A \cos{(m_A t - \vec{k} \cdot \vec{x} + \delta_\tau(t))}$ at location $\vec{x}$, the oscillating force on a mirror is given by $\vec{F}=F_0 \vec{e}_A \sin{(m_A t - \vec{k} \cdot \vec{x} + \delta_\tau(t))}$. The phase factor $\delta_\tau(t)$ varies on the coherence timescale of dark matter, given by $\tau = 2 \pi / (m_A v^2)$, where $v \simeq 10^{-3}$ represents the local velocity of dark matter. For a mirror with mass $M$ and $B-L$ charge $q_{B-L}$, the force amplitude is
\begin{equation}
 F_0 = \epsilon_{B-L} e q_{B-L} m_A A_0 \simeq \epsilon_{B-L} q_{B-L} \times 6.1 \times 10^{-16} \unit{N} \simeq \epsilon_{B-L} M \times 1.8 \times 10^{-11} \unit{m/s^2} .
\end{equation}
Here, $\epsilon_{B-L}$ is the gauge coupling constant normalized to the electromagnetic coupling, $\vec{e}_A$ is the unit vector parallel to $\vec{A}$, and $k = m_A v$. We assumed that the field energy density equals the local dark matter density, $\rho_a = m_A^2 A_0^2 / 2 \simeq 0.4 \unit{GeV/cm}^3$. The amplitude of the oscillating displacement caused by this force is proportional to $g_{B-L}/M$, differing between materials. For example, for sapphire, which is used for the test mass mirrors of KAGRA, $g_{B-L}/M \simeq 0.510/m_{\rm n}$, while for fused silica, used for other mirrors of KAGRA, $g_{B-L}/M \simeq 0.501/m_{\rm n}$.

Gravitational wave detectors are highly sensitive to differential arm length (DARM) changes in two perpendicular arm cavities, as gravitational waves induce such changes. However, displacements caused by the vector field are largely common across the four test mass mirrors, as they are made of the same material, resulting in the cancellation of most vector dark matter signals. The DARM channel's sensitivity to vector dark matter comes from residual effects due to oscillation phase differences and finite light-travel time~\cite{MorisakiFinite}. Despite this cancellation, LIGO and Virgo have set some of the most stringent constraints, as shown in Fig.~\ref{fig:sensitivity} (left).

KAGRA is the only gravitational wave detector with mirrors made of different materials. Measuring distance changes between sapphire test masses and fused silica auxiliary mirrors enhances sensitivity to $B-L$ dark matter~\cite{MichimuraVector}. Auxiliary length channels, such as the differential Michelson interferometer length (MICH), power recycling cavity length (PRCL), and signal recycling cavity length (SRCL), can be used to search for the signal. Once the detector reaches its designed sensitivity, it will surpass equivalence principle test limits, as shown in Fig.~\ref{fig:sensitivity} (left).

Using data from KAGRA's first joint observing run O3GK with the GEO600 detector in 2020, we set upper limits on the $B-L$ gauge coupling at the $10^{-19}$ level in the mass range $10^{-13} \unit{eV} \lesssim m_A \lesssim 10^{-12} \unit{eV}$~\cite{O3GKVector}. This analysis involved developing a new pipeline to search for oscillatory length changes while carefully accounting for the stochastic nature of ultralight dark matter signals~\cite{NakatsukaStochastic}. Since KAGRA had not yet reached its planned sensitivity during the O3GK run, our limits remain several orders of magnitude less stringent than those set by previous experiments. Nevertheless, this study highlights the potential of using auxiliary length channels in gravitational wave detectors for astrophysical observations, which were previously overlooked as useful for astrophysical studies. With dedicated noise reduction in these channels during future runs, previously unexplored regions could be probed.

\section{Polarization optics for axion dark matter search implemented in KAGRA}
Axions and axion-like particles are prominent dark matter candidates, with extensive efforts to search for their signatures through diverse experiments and astrophysical studies. Recent measurements of isotropic cosmic birefringence from cosmic microwave background polarization data have gained attention as a potential hint of axion physics, as axions exploit parity-violating interactions with photons. Cosmic birefringence induced by dark energy may also be linked to the coupling of axion dark matter to photons in a two-axion model~\cite{ObataTwoAxion}.

The axion-photon interaction, characterized by the coupling constant $\gagamma$, induces a phase velocity difference between left- and right-handed circular polarizations of light. Under the background axion field $a(t)=a_0 \cos{(m_a t + \delta_\tau(t))}$ with a mass $m_a$, the phase velocity difference $\delta c = c_{\rm L}-c_{\rm R} =\delta c_0 \sin{(m_a t+\delta_\tau(t))}$ for light with wavelength of $\lambda_{\rm l} = 2 \pi / k_{\rm l}$ is given by
\begin{equation}
 \delta c_0 = \frac{\gagamma a_0 m_a}{k_{\rm l}} \simeq 2.1 \times 10^{-24} \left( \frac{\lambda_{\rm l}}{1064 \unit{nm}} \right) \left( \frac{\gagamma}{10^{-12} \unit{GeV^{-1}}} \right) .
\end{equation}
This phase velocity difference induces an oscillating polarization rotation with an amplitude given by
\begin{equation}
 \delta \beta_0 = \frac{\gagamma a_0 m_a L}{2\sqrt{2}} \simeq 1.3 \times 10^{-14} \unit{rad} \left( \frac{L}{3 \unit{km}} \right) \left( \frac{\gagamma}{10^{-12} \unit{GeV^{-1}}} \right) ,
\end{equation}
for light propagation over a distance $L$~\cite{ADBCFirst}. Such effects can be investigated using table-top ring cavity experiments like DANCE~\cite{DANCE}, or linear cavities in laser interferometric gravitational wave detectors via the ADAM-GD scheme~\cite{ADAM-GD,ADAM-GD2}.

Figure~\ref{fig:sensitivity} (right) shows the projected sensitivity of the KAGRA detector using the ADAM-GD scheme. Since the arm cavities are over-coupled, the search using the cavity reflected beams at the anti-symmetric port generally offers better sensitivity~\cite{ADAM-GD}. However, the search using the arm cavity transmitted beams provides better sensitivity at lower masses. This difference arises because the light at the transmission ports travels through the cavity an odd number of times, and the polarization rotation effect is not canceled by polarization flipping due to mirror reflections~\cite{ADAM-GD2}. In both cases, sensitivity peaks at $m_a = \pi (2N - 1) / L_{\rm cav}$, where $L_{\rm cav} = 3$~km is the arm cavity length and $N \in \mathbb{N}$. To search for the polarization rotation of laser beams, we installed polarization optics at the transmission ports of the two arm cavities in 2021 and are now ready to collect the first axion data during the upcoming O4c observing run in 2025.

\section*{Acknowledgments}
This work was supported by JSPS KAKENHI Grant Nos.~19K14702, 20H05639, 20H05850, 20H05854, 20H05859, 23H00110, 23H04891, 23H04893, 23K03408 and 24K00640, and by JST PRESTO Grant No. JPMJPR200B.

\end{document}